\documentclass{llncs}

\usepackage{graphicx}
\usepackage{cite}
\usepackage{url}

\begin{document}

\title{Security Analytics of Network Flow Data of IoT and Mobile Devices \\(Work-in-progress)}
%If Title is too long, use \titlerunning
%\titlerunning{Short Title}

%Multiple institutes are typeset as follows:
\author{Ashish Kundu\inst{1}
		\and Chinmay Kundu\inst{2}
		\and Karan K. Budhraja\inst{3}}
%If there are too many authors, use \authorrunning
%\authorrunning{First Author et al.}

\institute{
IBM Thomas J. Watson Research Center, Yorktown Heights, NY, USA\\
\email{akundu@us.ibm.com}\and
KIIT University, Bhubaneswar, India\\
\email{ckkundu@gmail.com}\and
University of Maryland, Baltimore County, MD, USA\\
\email{karanb1@umbc.edu}
}
			
\maketitle

\begin{abstract}
Given that security threats and privacy breaches are commonplace today, it is an important problem for one to know whether their device(s) are in a "good state of security", or is there a set of high-risk vulnerabilities that need to be addressed. In this paper, we address this simple yet challenging problem. Instead of gaining white-box access to the device, which offers privacy and other system issues, we rely on network logs and events collected offline as well as in realtime. Our approach is to apply analytics and machine learning for network security analysis as well as analysis of the security of the overall device - apps, the OS and the data on the device. We propose techniques based on analytics in order to determine sensitivity of the device, vulnerability rank of apps and of the device, degree of compromise of apps and of the device, as well as how to define the state of security of the device based on these metrics. Such metrics can be used further in machine learning models in order to predict the users of the device of high risk states, and how to avoid such risks.

\end{abstract}

%\begin{keywords}
%keyword1, keyword2
%\end{keywords}

\section{Introduction}
\label{section:introduction}

\begin{figure}
\centering
\includegraphics[width=\textwidth]{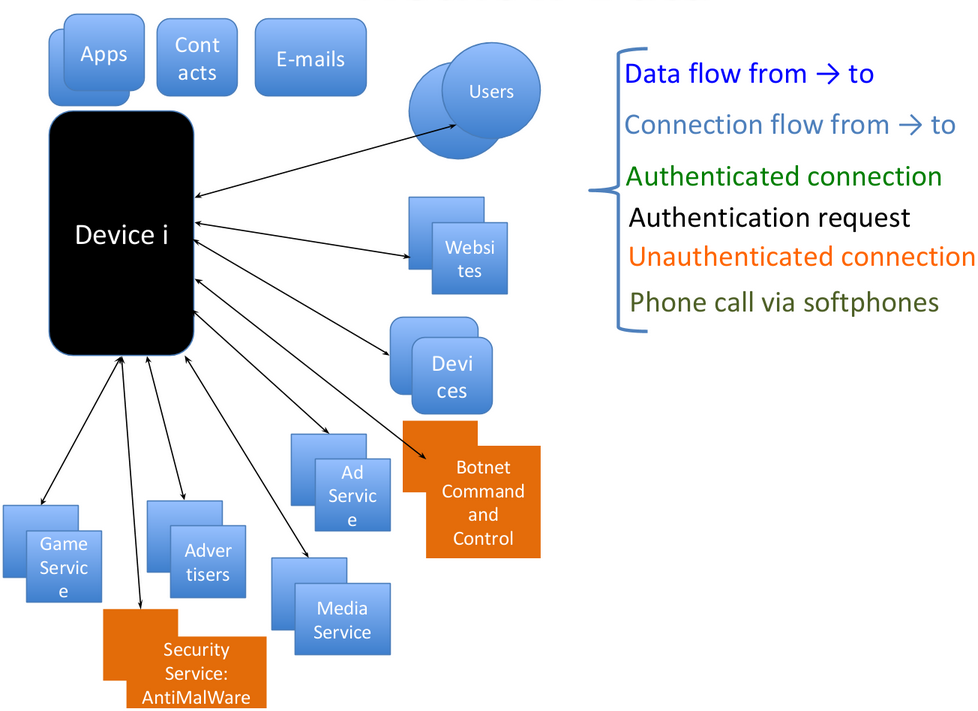}
\caption{Summary of network interactions for a network device.}
\label{fig:network_interactions}
\end{figure}

Network flow data may be categorized as encrypted and unencrypted traffic. Encrypted traffic allows for transmission specific information such as TCP/IP headers, session information, and details of the cipher suite being used. Unencrypted traffic allows for transmission specific information such as URL tags, the context, version and name of an application, DUID and UID, location, the name and type of the device, and details of the operating system being used. The various network interactions of a device are summarized in Figure \ref{fig:network_interactions}.

%\section{Related Work}
%\label{section:related_work}

%todo

\section{Inferences from Network Logs and Events}
\label{section:problem_definition}

The problem addressed by this work is presented using the generation of $4$ inferences.

\subsection{Inference $1$}

The first problem is that of inference of details of a network device and the communication associated with it. These details may be smummarized as follows.

\begin{enumerate}
\item The type of device and its associated software versions.
\item The encryption being used in the traffic associated with the device e.g., DHCP and URL type (e.g., m.google.com, appstore, and other standard IP addresses and URLs being used by current iPhone and Android devices).
\item The information embedded in network packets.
\item The location of a device, observable from its IP address (HTTPS) and HTTP headers.
\item The communication network associated with the device (e.g., services and IP addresses exracted from TCP/IP header information).
\item The frequency of usage of the device and the duration of each session.
\item The advertisements catered to by the device.
\item The categories of applications that are used on the device (e.g., work, games, utility, phone, media, social networks, push type, and pull type).

\end{enumerate}

Application specific information may also be identified. This includes the server addresses used by the application to communicate to a third-party server, if any. This also includes distribution information such as the application version and vendor (may be obtained from application hosting website). Further details include those of device users: the number of users and their identities. Miscellaneous details about the application include the advertisement providers (e.g., Google advertisements) that it communicates wth and the means of communication (e.g., HTTP, HTTPS).

This inference also involves the accumulation of network-based statistics. These include the statistics of SSL/TLS used and the various security services used across the network (e.g., communication with Symantec servers). An additional device type may also be obtained for TPM-bsaed services. The statistics also include identification of the type of payload used in encrypted traffic.

\subsection{Inference $2$}

This inference focuses on the communication between devices and applications across the network. The inference involves the generation of user and application profiles corresponding to network data transactions. It also involves identification of the distribution network used by advertisements. These observations can collectively be used to identify a network map of device vulnerability (e.g., the potential behavior of a device with respect to malware propogation across the network, or the observation of botnet components in the network).

\subsection{Inference $3$}

This inference pertains to the compromise of a device and its associated software. An example may be a focus on identification of rootkits in devices. These may be identified by observation such as the way in which the device behaves on the network, the type of operating system and applications installed, the web addresses that the device communicates to, and a list of potential vulnerabilities.

Hypotheses about an application may be formulated by comparing application behavior with signatures of compromised behavior. This may be performed for identification of applications which are either compromised or exhibit potential of being compromised in the future. This may also identify applications which are currently not compromised but were compromised in the past.

This information allows for the identification of a security lifecycle associated with a network device, where a device moves between protected and compromised states. Arcs in such a state diagram may correspond to details about how the device was cleaned of the compromise e.g., the use of a new installation of the application. It may additionally encode information about the data transmitted during the compromised state.

Analysis of this state diagram allows for identification of applications as malware or re-packaged applications. Such applications exhibit the threat of leaking confidential or private data, user behavior patterns, and sensory data. They may also transmit information related to communication channels used by the device and their associated advertisements and security measures (such as CAPTCHA). The identification of such threats is useful to avoid the compromise of host security and to avoid the propagation of malware.

Finally, such information may be used for forensic and security breach analysis. This involves identification of the types of breaches that may have occurred. This comprises of observation of whether the device is compromised and whether it exhibits anomalous behavior. This translates to servers with which the device communicates and their frequency and duration. Analysis may also involve the observation of an increase in advertisements or the vulnerability of the device. Note that the potential cost of such security breaches is difficult to estimate without the availability of whitebox information for the network. Such availability includes the knowledge of details about the compromise (e.g., the cause and duration of the compromise, and the network actions that were executed by the attacker during that period).

\subsection{Inference $4$}

The final inference to be generated is the energy usage profile of the user. Energy usage allows for a secondary estimate of data transaction by applications (independent applications and the use of web pages). Note that such inference is more useful when the operating system or application being used exhibits adaptive energy consumption (optimizing for lower energy consumption). Since such behavior is integrated in most modern-day software, an energy usage profile is capable of functioning as a secondary source of information on data transactions.

\section{Proposed Method}
\label{section:proposed_method}

\begin{figure}
\centering
\includegraphics[width=\textwidth]{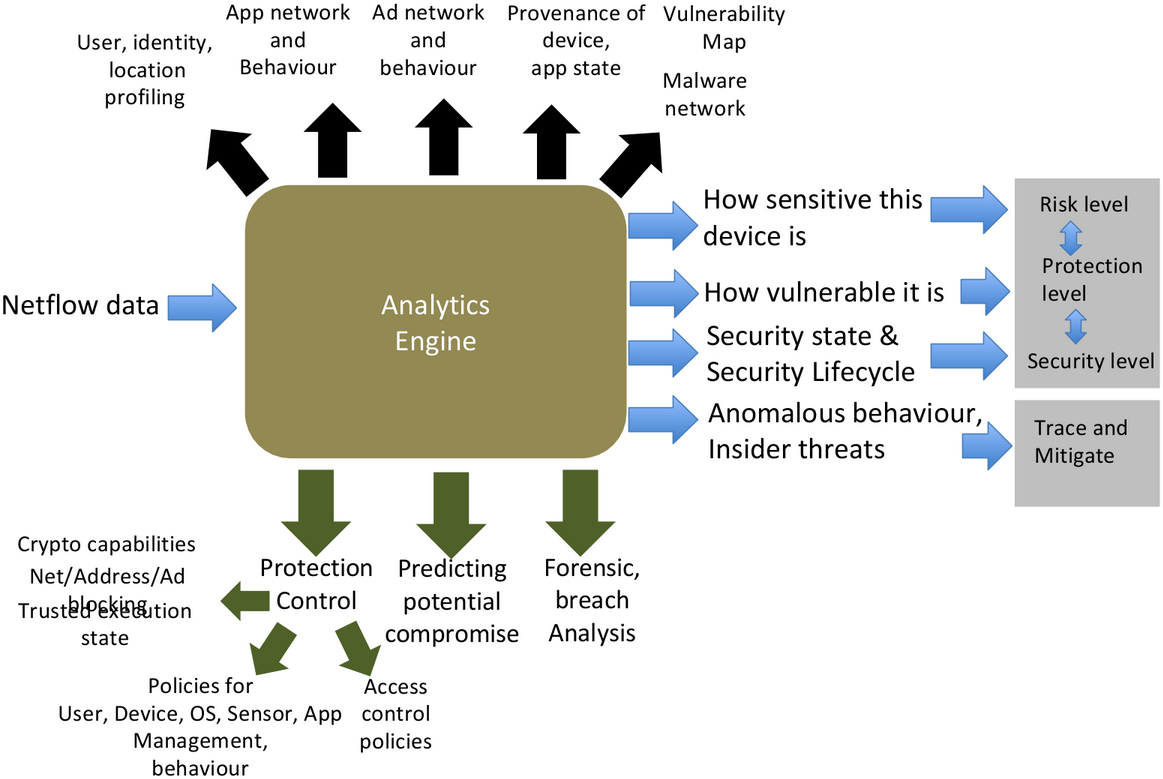}
\caption{Using network flow data.}
\label{fig:using_network_flow_data}
\end{figure}

The various uutilities of network flow data are summarized in Figure \ref{fig:using_network_flow_data}. The proposed method then is divided into sections based on the different aspects of analysis of network data.

\subsection{Sensitivity Analysis}

Sensitivity of a device is defined as the importance of data that the device is used with. Specifically, this may be defined as the extent of the device data being personal to the user. The analysis of sensitivity of a device is dependent on the following factors. Note that the analysis is recusive with respect to the sensitivity of the components involved.

\begin{enumerate}
\item The sensitivity of the data stored, generated and deleted dynamically (e.g., personal data, passwords and cookies).
\item The sensitivity of the applications used, and data stored by them. Note that the sensitivity of an application may depend on its age.
\item The sensitivity of the other devices and web sites that the user connects to, using this device.
\item The sensitivity of the individuals (other users) that the device is used to contact (e.g., via phone calls, messaging, calls).
\item The sensitivity of the connections and communications associated with the device (e.g., frequency, time, and identity).
\item The amount of data transferred.
\end{enumerate}

\subsection{Sensitivity Rank}
\label{subsection:sensitivity_rank}

The sensitivity rank of a device implies the risk level and protection level associated with that device, with respect to sensitivity. Because of the dynamic data and applications associated with the device, sensitivity rank is a dynamic value. For example, a wallet is highly sensitive if it contains more currency, credit cards, personal info, or a combination of these.

Sensitivity rank is computed by building a sensitivity graph. The nodes for this graph correspond to the entities involves i.e., the devices. The edges for this graphs correspond to communication from one entity to another. Sensitivity rank may then be computed as a probabilistic value, similar to PageRank \cite{page1999pagerank}. We therefore define $S(x)$ as the sensitivity rank function for entity $x$. Similar to PageRank, we have $S(x) \in (0,1)$.

The following discusses the sensitivity relation between two entities. The sensitivity rank, $S(j)$ of an entity $j$ that entity $i$ communicates with, contributes a weighted value to the sensitivity rank $S(i)$ of device $i$. A weight function (for an edge connnecting entity $i$ and entity $j$) is defined such that the weight function $W(i,j)$ depends on sensitivity related information such as the frequency of communication, usage, amount of data transferred, how far in the timeline the communication was carried out, age of apps, authenticated or unauthenticated connection. $W(i,j)$ therefore represents weight assigned to the \textit{relative rank} that entity $j$ contributes to $S(i)$.

A high-level recurrence formulation is then presented in Equation \ref{eq:sensitivity_rank}.

\begin{equation}
S(i) = \sum(W(i,j)S(j)) + S(D(i))
\label{eq:sensitivity_rank}
\end{equation}

Where a data function, $D(i)$, is included for increased precision. Specifically, $D(i)$ represents the data stored at entity $i$. This includes the data stored across applications. $D(i)$ is decomposed in Equation \ref{eq:d_i}. $HD(i)$ represents the data generated by and stored at entity $i$ in the past. $CD(i)$ represents the current data generated by and stored at entity $i$.

\begin{equation}
D(i) = HD(i) + CD(i)
\label{eq:d_i}
\end{equation}

Precise computation of sensitivity rank is also dependent on data sources other than those which provide TCP/IP information. This may be latent information learned from network flow data. Note that many applications are identifiable because they do not use TLS. They may also be identified from from the advertisements clicked on by the user, or presented as a part of the application.

\subsection{Vulnerability Analysis}

Vulnerability of a device is defined as the ease with which a device may be compromised. The analysis of volnerability of a device is dependent on the following factors. Similar to sensitivuty, the analysis is recusive with respect to the vulnerability of the components involved.

\begin{enumerate}
\item The vulnerability of the applications installed on the device.
\item The vulnerability of the operating system being used by the device.
\item The possibility of vulnerability propagation, i.e. the possibility of a data transaction path existing between a compromised website and vulnerable application on the device.
\item The advertisements associated with applications. This includes the sources of the advertisements and the scripts that they may execute.
\item The information that is transmissed by applications. This comprises of the following.
\subitem Periodic notifications associated with the application. These include push notifications.
\subitem The fetching of advertisements and actions associated with clicking on an adversitement.
\subitem The transmission of sensor-based information. This is significant when considering the possibility of password cracking by the use of sensor readings.
\subitem The solving of CAPTCHAs \cite{von2003captcha} associated with the application.
\end{enumerate}

\subsection{Vulnerability Rank}
\label{subsection:vulnerability_rank}

Similar to sensitivity rank, the computation of vulnerability rank requires the construction of a graph. A node $(x,V(x))$ in the graph represents entity $x$ with vulnerability $V(x)$, where $V(x)$ is the vulnerability function. An edge from entity $x$ to entity $y$ exists if entity $x$ is compromised by exploiting the vulnerability $V(x)$ and vulnerability $V(y)$. The probability of these vulnerabilities being exploited is reprented by the join probability $p((x,V(x)),(y,V(y))) \in (0,1)$.

The graph is constructed from the known vulnerabilities of different components involved e.g., the operating system, applications, vulnerable websites and services, and advertisements. For an advertisement, the graph may consider how vulnerable or malicious the advertisement nework is. The vulnerability of an application depends on the vulnerability of the advertisements that it receives, the source of advertisements, and scripts that are executed thereof. An example of using the graph may be inspection of the existence of a path from a compromised entity to the user's device (paths with $p>0$. Note that an entity in this graph may be a device, an application, an advertisement, an operating system, or even hardware and firmware components.

As in sensitivity rank, vulnerability rank may then be computed as a probabilistic value, similar to PageRank \cite{page1999pagerank}. We therefore define $V(x)$ as the vulnerability rank function for entity $x$. Similar to PageRank, we have $V(x) \in (0,1)$.

The following discusses the vulnerability relation between two entities. The vulnerability rank $V(j)$ of an entity $j$ that entity $i$ interacts with, contributes a weighted value to the vulnerability rank $V(i)$ of entity $i$. The weight function $W(i,j)$ depends on $p(x,y)$, the frequency of communication, usage of the entity, the amount of data used, the relative time of occurrence of the communication, and the age of applications associated with the entities. Note that the weight value $W(i,j) \in (0,1)$.

A high-level recurrence formulation (for the interaction between entity $i$ and entity $j$) is then presented in Equation \ref{eq:vulnerability_rank}. This formulation includes Denial of Service (DoS) \cite{needham1993denial} as a potential threat. Even though an entity may not have any vulnerability (which is highly improbable), $LV(i)$ does not influence the first term in Equation \ref{eq:vulnerability_rank}, that is due to the remote entities that entity $i$ interacts with.

\begin{equation}
V(i) = \sum(W(i,j)V(j)) + IV(i) + LV(i)
\label{eq:vulnerability_rank}
\end{equation}

Where $IV(i)$ and $LV(i)$ are additional vulnerability functions used for increased precision. $IV(i)$ represents vulnerability due to insiders i.e., the probability that an insider (authorized user with respect to entity $i$) would become a proponent of compromise of entity $i$. $LV(i)$ represents the local vulnerability of entity $i$ i.e., the probability that compoenents within entity $i$ can be expoited to produce a compromise.

\subsection{Degree of Compromise}
\label{subsection:degree_of_compromise}

The degree of compromise is a composite metric formulated using sensitivity and vulnerability. Degree of compromise, $DC(i)$ of a component $i$, is based on the following observations in network data.

\begin{enumerate}
\item The components and applications specific to a device, that have been compromised.
\item The probability $p(i)$ that the device is compromised based on vulnerability rank $V(i)$ and the network behaviour of the entity $i$.
\item The criticality of teh compromise of entity $i$, based on the corresponding sensitivity rank $S(i)$.
\end{enumerate}

A high-level recurrence formulation is then presented in Equation \ref{eq:degree_of_compromise}. $j$ is the component (such as the operating system or application) on the device that has a vulnerability rank $V(j) > 0$.

\begin{equation}
DC(i) = p(i)*S(i)*Sum(DC(j))
\label{eq:degree_of_compromise}
\end{equation}

\subsection{Security State Analysis}

\begin{figure}
\centering
\includegraphics[width=2.5in]{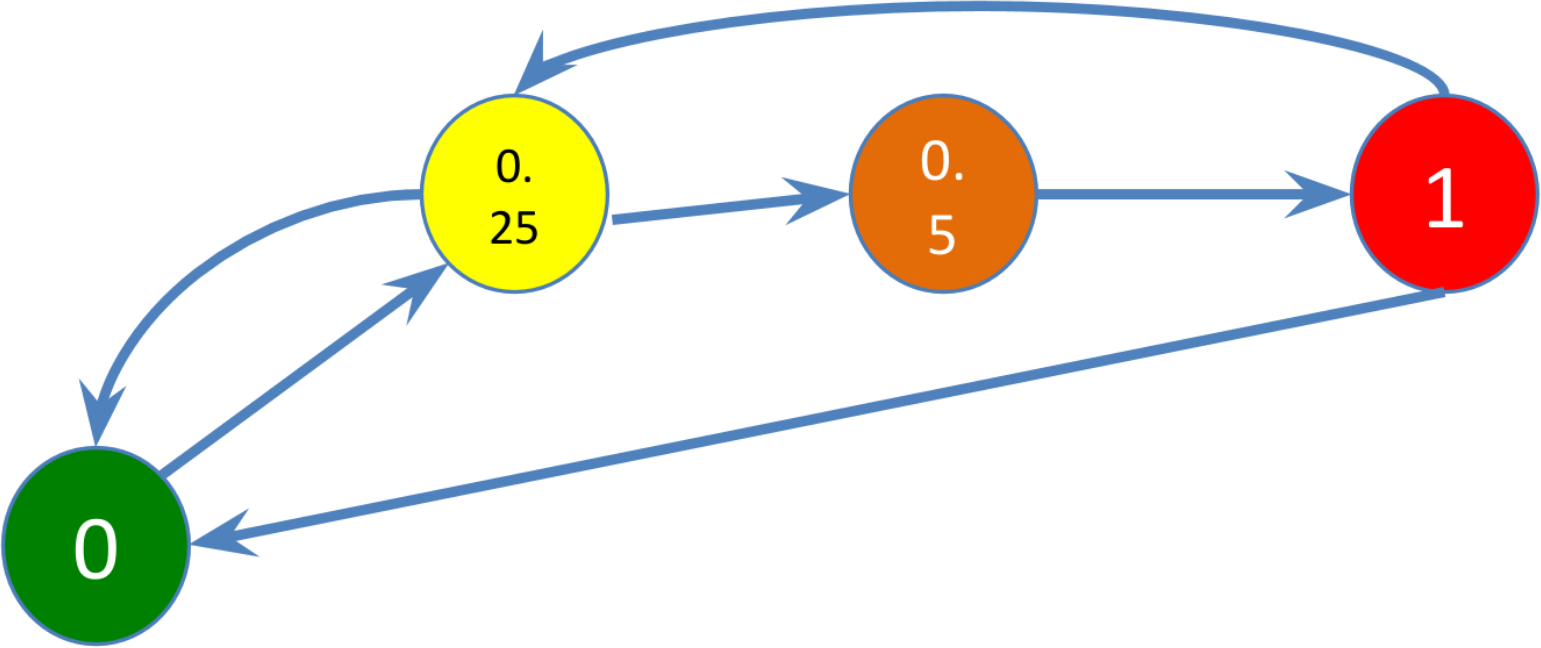}
\caption{Security state diagram for a specific device. The diagram and values associated with nodes are dynamic with respect to the applications installed in the device and the network data transactions thereof. The values on nodes represent the degree of compromise.}
\label{fig:state_diagram}
\end{figure}

\begin{figure}
\centering
\includegraphics[width=2.5in]{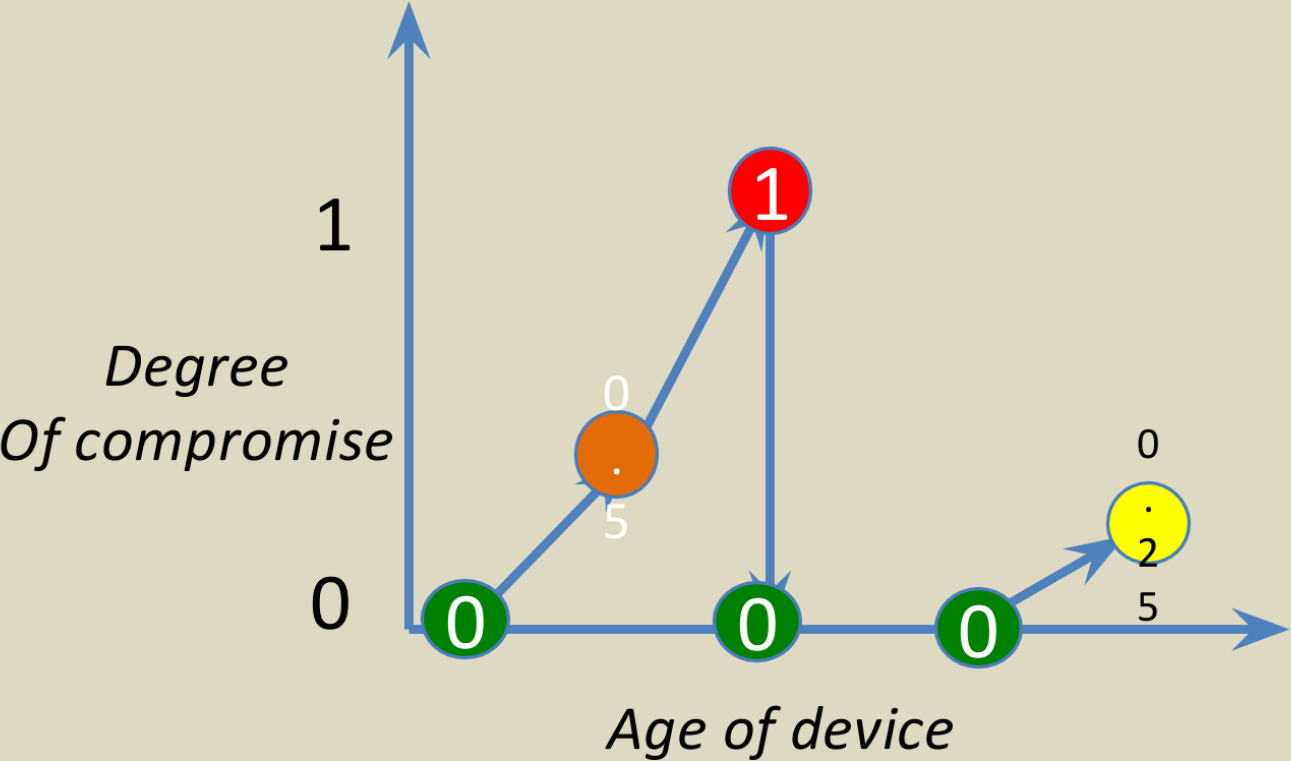}
\caption{Security state transitions.}
\label{fig:state_transition}
\end{figure}

The security state of a device can be determined from network data. The formulation of such state dynamics is useful for applications such as risk analysis, taking protective actions in case of a compromise of security, and forensic and security breach analysis. The security state diagram of a device is summarized in Figure \ref{fig:state_diagram}. The transitions between different security states over time are summarized in Figure \ref{fig:state_transition}. The security state of a network device may be determined by the following.

\begin{enumerate}
\item The observation of anomalous behavior with respect to the device.
\item The access of security services (e.g., Symantec servers) by the device.
\item The strength of the cipher suite associated with the device.
\item The protocol used by the device (e.g., HTTP, HTTPS, SRTP, and RTP).
\end{enumerate}

\subsection{Protection Based on Network Flow Data Analysis}

At a given time, the protection level required to prevent the risk can be associated with the quantification of the risk (involving the computation of $S(i)$, $V(i)$ and $DC(i)$). The details of these computations are presented in Section \ref{subsection:sensitivity_rank}, Section \ref{subsection:vulnerability_rank} and Section \ref{subsection:degree_of_compromise}. This quantity can then be limited to a threshold. Appropriate actions can then be incorporated across the network. Examples of the different types of actions that may be executed across the network are enumerated as follows.

\begin{enumerate}
\item Strengthen or weaken access control and authorization policies with respect to different network components.
\item Restart a network component. The device may then enter a trusted execution mode, e.g., for financial transactions. This may be required for a network component $i$ that observes $DC(i) > 0$. Further, all future data transactions may be monitored for the given device.
\item Notification of the current security state of a user, and appropriate actions that may be taken.
\item The enforcement of user behavior policies. For example, a user may not be allowed to open a sensitive website such as www.chase.com after it has clicked on an advertisement from low-sensitivity application such as a game (e.g., AngryBirds).
\item The enforcement of mixed user behavior policies with system-driven control.
\item The enabling or disabliing of applications, features and sensors across the network based on updated policies.
\item Initiation of backup of device data and the removal of all sensitive data and applications from the device.
\item The blocking of third-party application synchronization (with other devices) for vulnerable applications.
\item The lockdown of network communication at various levels of granularities across the network.
\item Alternative defense mechanisms for the network, in the case where the policy engine governing the network has been compromised.
\item Disabling a device by draining or removal of all associated energy sources. This may be enabled by a remote protection unit that sends targeted scripts via advertisements and web pages to drain energy. For example, if the device is being used as a bot, and the network infrastructure is beyond control by local protection.
\end{enumerate}

Risk analysis may also be translated to long-term actions such as the following.

\begin{enumerate}
\item The identification of requirements for software patches required across different devices and types of patches, and their scheduling. This may be based on the security state analysis of a device.
\item The engineering of applications and software. This may depend on the programming models to be enforced, analysis that is required to be incorporated with the program (such as sensitivity and vulnerability).
\item The devising of a methodology by which applications and software are requred to support APIs for usage with trusted local and remote services. This enforeces dynamic protection and the ability of take appropriate defense actions.
\end{enumerate}

\subsection{Forensic Analysis}

Forensic analysis of network data involves the use of provenance data collected by the network. This data comprises of the following.

\begin{enumerate}
\item Information about the states of different network components
\item Information about the various ranks (such as sensitivity and vulnerability) and degrees of different network components.
\item Other latent information that may be inferred from data transmission across the network.
\end{enumerate}

Such analysis is useful for automated auditing of the behavior of a device, users and applications that are present in the network. An alert may be issued by a background process that continuously checks whether the degree of compromise crosses a threshold value. This process may then also identify the following information.

\begin{enumerate}
\item An analysis of any breaches that may have occurred.
\item The methodology that was adopted to compromise a given device.
\item The lack of protections in context of netwrk security that may be incorporated to avoid such compromise in the future.
\end{enumerate}

Alternatively, forensic analysis may be used to analyze the potential cost of a security breach. This may be based on the following.

\begin{enumerate}
\item The end-points that the device is communicating with at the time of compromise.
\item Details about the communication, such as its duration and the frequency of such communication.
\item Categorizaton of the communication to reflect severity of the potential breach. Examples of categories of communication include monetary, political, social, and private.
\end{enumerate}

\section{Machine Learning}
\label{section:ml}

We develop a Spark-based machine learning stack for implementation of classification of risk levels of apps and of a device. We plan to develop an SVM-based system, following which we plan to develop a neural-network based model and compare their accuracy versus efficiency in predicting risks as well as classification of apps in the risk lattice. We plan to apply the risk classification and risk prediction for multiple devices together that are in a geolocation or in a network.

\section{Conclusions}
\label{section:conclusion}

In this paper, we discussed the problem of determining the state of security of a device -- mobile or IoT, using big data analytics and machine learning on network logs and events of such devices. We further outlined a set of steps towards remediation of such issues.

\bibliographystyle{splncs03}
\bibliography{nfdsecurity}

\begin{thebibliography}{1}
\providecommand{\url}[1]{\texttt{#1}}
\providecommand{\urlprefix}{URL }

\bibitem{needham1993denial}
Needham, R.M.: Denial of service. In: Proceedings of the 1st ACM Conference on
  Computer and Communications Security. pp. 151--153. ACM (1993)

\bibitem{page1999pagerank}
Page, L., Brin, S., Motwani, R., Winograd, T.: The pagerank citation ranking:
  Bringing order to the web. Tech. rep., Stanford InfoLab (1999)

\bibitem{von2003captcha}
Von~Ahn, L., Blum, M., Hopper, N.J., Langford, J.: Captcha: Using hard ai
  problems for security. In: International Conference on the Theory and
  Applications of Cryptographic Techniques. pp. 294--311. Springer (2003)

\end{thebibliography}

\end{document}